\documentclass[aps,pra,floatfix,nofootinbib,longbibliograpphy]{revtex4-2}
\usepackage{dcolumn}
\usepackage{bm}
\usepackage{graphicx,subfigure,xcolor}
\usepackage{braket}
\usepackage[normalem]{ulem}
\usepackage{balance}
\usepackage{comment}
\usepackage{xcolor}
\usepackage{amsmath}
\usepackage{amssymb}
\usepackage{eucal}
\usepackage{mathrsfs}
\usepackage{amsthm}
\usepackage{epstopdf}
\usepackage{textcomp}
\usepackage{braket}
\usepackage[utf8]{inputenc}
\usepackage[T1]{fontenc}
\usepackage{xcolor}
\usepackage{soul}

\begin{document}

\title{Spectroscopy of alkali atoms in solid matrices of rare gases: experimental results and theoretical analysis}

\newcommand{\wo}{\omega_0}
\newcommand{\wk}{{\omega_k}}
\newcommand{\wkp}{\omega_{k'}}

\newcommand{\ewo}{e^{i\omega_0 t}}
\newcommand{\ewom}{e^{-i\omega_0 t}}
\newcommand{\ewop}{e^{i\omega_0 t'}}
\newcommand{\ewomp}{e^{-i\omega_0 t'}}

\def\bk{{\bf k}}
\def\bkp{{\bf k}'}
\def\bkj{{\bf k}j}
\def\br{{\mathbf r}}
\def\bR{{\mathbf R}}
\def\wp{{\omega_p}}
\def\bfm{{\bf f}}

\newcommand{\bmu}{\boldsymbol\mu}
\newcommand{\vac}{\vert vac\rangle}
\newcommand{\ewk}{e^{i\omega_k t}}
\newcommand{\ewkm}{e^{-i\omega_k t}}
\newcommand{\ewkp}{e^{i\omega_k' t}}
\newcommand{\ewkmp}{e^{-i\omega_k' t}}
\newcommand{\ekr}{e^{i{\bf k}\cdot{\bf r}}}
\newcommand{\ekrm}{e^{-i{\bf k}\cdot{\bf r}}}
\newcommand{\ekri}{e^{i{\bf k}\cdot{\bf r}_i}}
\newcommand{\ekrmi}{e^{-i{\bf k}\cdot{\bf r}_i}}
\newcommand{\ekrA}{e^{i{\bf k}\cdot{\bf r}_A}}
\newcommand{\ekrmA}{e^{-i{\bf k}\cdot{\bf r}_A}}
\newcommand{\ekrAp}{e^{i{\bf k}'\cdot{\bf r}_A}}
\newcommand{\ekrmAp}{e^{-i{\bf k}'\cdot{\bf r}_A}}
\newcommand{\skj}{\sum_{{\bf k}j}}
\newcommand{\skjp}{\sum_{{\bf k'}j'}}
\newcommand{\akjd}{a^{\dag}_{{\bf k}j}}
\newcommand{\akj}{a_{{\bf k}j}}
\newcommand{\akjdp}{a^{\dag}_{{\bf k'}j'}}
\newcommand{\akjp}{a_{{\bf k'}j'}}
\newcommand{\ekj}{\hat{{\bf e}}_{{\bf k}j}}
\newcommand{\ekjp}{{\bf e}_{{\bf k'}j'}}

\author{Caterina Braggio$^{1}$}
\author{Roberto Calabrese$^{2}$}
\author{Giovanni Carugno$^{1}$}
\author{Giuseppe Fiscelli$^{3}$}
\author{Marco Guarise$^{2}$}
\author{Alen Khanbekyan$^{2}$}
\author{Antonio Noto$^{3}$}
\author{Roberto Passante$^{3,4}$}
\author{Lucia Rizzuto$^{3,4}$}
\author{Giuseppe Ruoso$^{5}$}
\author{Luca Tomassetti$^{2}$}

\affiliation{$^1$ Dipartimento di Fisica e Astronomia ''G. Galilei",
Universit\`{a} di Padova and INFN Padova, Via F. Marzolo 8, I-35131
Padova, Italy}
\affiliation{$^2$ Dipartimento di Fisica e Scienze della Terra,
Universit\`{a} di Ferrara and INFN Ferrara, Via G. Saragat 1, I-44122
Ferrara, Italy}
\affiliation{$^3$ Dipartimento di Fisica e Chimica - Emilio Segr\`{e},
Universit\`{a} degli Studi di Palermo, Via Archirafi 36, I-90123 Palermo,
Italy}
\affiliation{$^4$ INFN, Laboratori Nazionali del Sud, I-95123 Catania, Italy}
\affiliation{$^5$ INFN, Laboratori Nazionali di Legnaro, Viale
dell'Universit\`{a} 1, I-35020 Legnaro, Italy}

\begin{abstract}
We present an experimental and theoretical investigation of the spectroscopy of dilute alkali atoms in a solid matrix of inert gases at cryogenic temperatures, specifically Rubidium atoms in a solid Argon or Neon matrix, and related aspects of the interaction energies between the alkali atoms and the atoms of the solid matrix. The system considered is relevant for matrix isolation spectroscopy, and it is at the basis of a recently proposed detector of cosmological axions, exploiting magnetic-type transitions between Zeeman sublevels of alkali atoms in a magnetic field, tuned to the axion mass, assumed in the meV range. Axions are one of the supposed constituents of the dark matter (DM) of the Universe. This kind of spectroscopy could be also relevant for the experimental search of new physics beyond the Standard Model, in particular the search of violations of time-reversal or parity-charge-conjugation (CP) symmetry. In order to efficiently resolve the axion-induced transition in alkali-doped solid matrices, it is necessary to reduce as much as possible the spectral linewidth of the electronic transitions involved. The theoretical investigation presented in this paper aims to estimate the order of magnitude of the inhomogeneous contribution to the linewidth due to the alkali--matrix interactions (Coulomb/exchange and dispersion), and to compare the theoretical results with our experimental measurements of spectra of dilute Rubidium atoms in Argon and Neon solid matrix. The comparison of the expected or measured spectral linewidths will be important for selecting the most appropriate combination of alkali atoms and matrix inert elements to be used in the proposed axion detection scheme. It is finally suggested that dilute Lithium atoms diffused in a cold parahydrogen solid matrix could be, overall, a good system upon which the proposed detector could be based.
\end{abstract}

\maketitle

\section{\label{sec:1}Introduction}

Matrix Isolation Spectroscopy (MIS) of alkali atoms in solid matrices of inert gases at cryogenic temperatures is a widely used spectroscopic technique in several fields of research \cite{Bondybey-Smith96,Momose-Shida98,Barnes-OrvilleThomas81}.
The atoms are trapped in fixed positions inside the solid matrix, and thus their mutual interactions are strongly reduced, as well as their interaction with the matrix atoms or molecules.
Thermal effects are also strongly reduced by the use of cryogenic temperatures (some Kelvin) \cite{Gerhardt-Sin12,Lambo-Buchachenko12,Kanagin-Regmi13}.
Some decades ago, MIS of atoms or molecules has been proposed for searching permanent electric dipole moments \cite{Pryor-Wilczek87,Arndt-Kanorski93}.
In more recent times, much theoretical research as well as new experimental proposals have been devoted to investigate new physics beyond the Standard Model (SM) exploiting atoms and molecules trapped through this technique \cite{Kozlov-Derevianko06,Vutha-Horbatsch18,Guarise-Braggio20,Santamaria-Braggio15}. In particular, searches for time-reversal or CP symmetry violations, electric dipole moment of standard-model particles, or DM candidates have been carried-out through MIS (for reviews on these subjects, see for example \cite{Safranova-Budker18,Cairncross-Ye19,DeMille-Doyle17,Andreev-Ang18}.

Cosmological axions are supposed to be one of the possible constituents of the DM \cite{Feng10}, whose composition is still unknown \cite{Bertone-Hooper18}.
Axions were initially introduced by Peccei and Quinn in order to solve the strong CP problem in quantum chromodynamics \cite{Peccei-Quinn77,Peccei-Quinn77a,Weinberg78,Wilczek78}, and are assumed to interact very weakly with ordinary matter. Successively Sikivie proposed that they could induce magnetic-type transitions between atomic Zeeman sub-levels, yielding possibility of their indirect detection in atomic structures \cite{Sikivie14}. It is thus of fundamental importance to envisage possible mechanisms, and relative experiments, to detect these particles. Several possibilities have been recently proposed; for a review, see for example \cite{Feng10,Rosenberg15,Graham-Irastorza15,Bertone-Tait18,Chadha-Day-Ellis22,Irastorza-Redondo18,Semertzidis-Youn22,Braggio-Carugno17}.

Recently, possible detection schemes for cosmological axions in the meV energy range, based on electron transitions between Zeeman-shifted levels of dilute alkali or oxygen atoms \cite{Santamaria-Braggio15}, or their ionization \cite{Guarise-Braggio20}, embedded in a solid matrix of inert gases at cryogenic temperatures \cite{Guarise-Braggio17}, have been proposed.

\begin{figure}[!htbp]
\centering\includegraphics[width=5 cm]{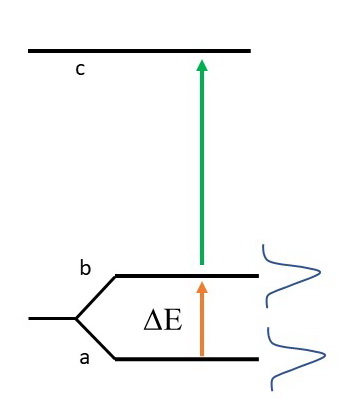}
\caption{The proposed scheme for the axion detection. The axion induces a transition between the two Zeeman sublevel $a$ and $b$, that can be tuned to the axion mass by the external magnetic field controlling the energy separation $\Delta E$. The absorption of the axion in then detected exploiting a laser field that induces an electronic transition from $b$ to the the conduction band of the matrix, whose edge is $c$, or, similarly, to an atomic excited level.}
\label{Fig1}
\end{figure}

The basic idea is illustrated in Fig. \ref{Fig1}: $a$ and $b$ denote two Zeeman sublevels of the guest atom, with energy $E_a$ and $E_b$, respectively; $\Delta E= E_b -E_a$ is their energy separation, that can be varied through the external magnetic field $B$. $\Delta E$ is tuned to the axion mass $m_a$, that is unknown, but supposed of the order of some meV. The absorption of the axion by the guest atom induces a transition from the Zeeman sublevel $a$ to the sublevel $b$.
The two Zeeman sublevels involved have the same electronic structure, so this transition does not induce any change of the electronic configuration, and possible linewidths due to electronic rearrangement or interatomic interactions or vibrational effects are absent.
Then, the atomic excitation can be detected exploiting a laser field that brings the atomic electron from $b$ to the conduction band of the solid matrix where the atom is embedded, or, with an alternative scheme, to an electronic excited state \cite{Santamaria-Braggio15}. Detection of the atomic excitation is thus a signature of the absorption of the axion. In order that this detection mechanism works efficiently it is necessary that the two sublevels $a$ and $b$ can be sharply resolved as much as possible: this requires that the linewidth of the atomic levels involved be smaller than their energy separation.
In the case of excitation to a higher discrete state of the atom, the linewidth of the excited level $c$ should be minimized too. It is therefore necessary making as small as possible all possible effects and interactions leading to the broadening of the spectral lines; this is particularly important in our case, due to the very small energy separation of the two Zeeman sublevels, assumed of the order of the meV.
Some specific properties of the guest atoms or molecules are required, such as enough sensitivity to the external magnetic field, spherical symmetry of the wavefunction in order to have a vanishing electric dipole moment, and easy accomodation in the solid matrix, in order to minimize its deformation. In \cite{Santamaria-Braggio15} an oxygen molecule, \textsuperscript{16}O\textsubscript{2}, was used; in this paper we will consider alkali atoms, initially Rubidium, from both the experimental and theoretical point of view, and then Lithium (only a theoretical analysis). Alkali atoms are very sensitive to magnetic fields, have a spherically symmetric structure, and Lithium in particular, can have a small size. In this work we present some preliminary experimental and theoretical results along this direction of research. Furthermore the properties of these atoms, and Rubidium in particular, are well know not only in free space but also when embedded into solid crystals.

The points mentioned above are essential for the proposed axion detection scheme, since the linewidth of the guest-atom energy levels involved in the transitions should be smaller than the separation of the two Zeeman sublevels, in order to be able to separate them. This is not an easy task to accomplish. For example, as we will see later on, our spectroscopy measurements of Rubidium atoms in a Argon matrix show that the linewidth in that case can be relatively larger.
Let us consider for example a peak of the fluorescence spectrum obtained from our measurements, shown in the next section, with $\lambda_c \simeq 1155$ nm for the peak wavelength and a width $\Delta \lambda \simeq 39$ nm. The energy width of the line is thus $\Delta E \simeq 3.6 \cdot 10^{-2}$ eV, that is quite larger than what is necessary for resolving the two Zeeman sublevels.
A main motivation of this work is to provide some hints for finding different schemes that significantly reduce the linewidth, for example different combinations of guest and matrix atoms or molecules, or different transition schemes. Very small optical linewidths have been recently found for thulium atoms trapped in argon and neon solid matrices \cite{Gaire-Raman19}, but they involve a magnetic-dipole transition, that is not a useful one for our proposed detection scheme.

Matrix Isolation Spectroscopy seems to be an appropriate experimental technique for this purpose, because it allows to control efficiently the interactions between the atoms and the environment, to reduce strongly
the mutual interaction between the atoms due to their large separation, as well as suppressing the
thermal broadening. It consists in trapping the alkali atoms in a solid matrix of inert materials at cryogenic temperatures, such as rare gases (RG) or inert molecules such as, for example, p-H\textsubscript{2}, O\textsubscript{2}, CH\textsubscript{4}.
Rb atoms trapped in a magneto optical trap could in principle be also used to detect the axions. However, one of the advantages of using the MIS is related to the high density of target atoms that can be probed with respect to spectroscopic studies in the gas phase. This fact is important also in the particle detection view since a high number of target atoms is necessary in order to maximize the interaction probability. Other axion detection schemes have been proposed, for example based on microwave cavities. The advantage of using a specific detection schemes depends also on the assumed (unkown) mass of the axion. In the axion mass range we are considering, larger than $ \sim 10^{-4}$ eV, the axion-electron interaction upon which our detection scheme is based is more efficient, while a scheme based on microwave cavities is more advantageous for smaller masses \cite{Semertzidis-Youn22}. Also, differently from other schemes exploiting for example microwave cavities
in the GHz region to detect DM axions, our approach can be easily tuned to a large energy range in the meV scale by varying the magnetic field, offering thus the possibility to investigate more sectors in the DM scenario in a complementary way.

In the cases we consider in this paper, possibly relevant for DM search,
the free atoms or molecules considered, rare gases, alkali atoms or parahydrogen, have not
electric dipole moments and their wavefunctions have spherical symmetry. Furthermore, the guest-atom-matrix interactions, mainly dispersion interactions, are very weak, and thus we expect that this property is maintained also when the atoms are inside the solid matrix.

In order to find out which is the best combination of guest atoms and solid-matrix atoms or molecules, minimizing the spectral broadening, it is important to
obtain reliable spectroscopic data for different combinations of guest atoms (alkali atoms, in this paper) and inert-gas cold matrices.
It is also of fundamental importance developing simple theoretical models to estimate the potential energy between the atoms or molecules, which can be easily used to compare different combinations of guest atoms and solid-crystal atoms or molecules. Their predictions can be then compared with available experimental data. Analysis of experimental data and theoretical predictions can also give important hints on how the guest atoms diffuse in the cold matrix in specific experimental setups and depending from the preparation of the doped cold crystal.
In fact, the potential energies between the alkali atoms and the matrix atoms or molecules, as we will show, can
give an indication of the inhomogeneous broadening of the alkali spectral lines, due to the atoms-matrix interaction energies, as well as indications on their spatial distribution inside the cold matrix.
The linewidth of the Zeeman-shifted energy levels should be the same for the unperturbed atoms, i.e. in the absence of the magnetic field, since their electronic wavefunction is the same.

We wish to mention that, at the criogenic temperatures attainable in our proposed detection scheme based on Matrix Isolation Spectroscopy, considering a 100 GHz frequency between the two Zeeman sublevels (that corresponds to a temperature of about 5 K), and tuned to an axion mass around $0.5$ meV, essentially all guest atoms are in their ground state; thus, the axion field is going to excite them from the ground state. A more detailed discussion on this point for the proposed detection scheme, as well as on the expected counting rate and signal to noise ratio, can be found in \cite{Sikivie14,Braggio-Carugno17}, in terms of the local axion energy density, the axion decay constant and other relevant parameters. The signal to noise ratio depends also on the lifetime of the Zeeman sublevels; for example, in a system recently considered, assuming a level lifetime of 1 ms, axions with mass larger than 80 GHz can be searched with a signal to noise ratio of 3 and with a significant count number in one hour observation time, provided the crystal is cooled below 100 mK \cite{Braggio-Carugno17}.

In this paper, in section \ref{sec:experimental} we first present experimental spectroscopic data of dilute Rubidium atoms in Argon and Neon cold matrices, with a particular attention to the lineshape of the emission spectra. Next, in section \ref{sec2} we propose a simple general theoretical model of the interaction potential energy between the alkali atoms and the inert atoms or molecules of the matrix, in terms of few data that can be found in the literature (theoretically calculated or experimental), specifically equilibrium distance, dissociation energy, van der Waals constant, and the curvature of the potential energy at the equilibrium distance.
The theoretical results are then used to estimate the order of magnitude of the broadening of the alkali spectral lines due to their interaction with the atoms or molecules of the cold matrix. This expected broadening is compared with our spectral experimental results. The results
can be usefully exploited also to estimate the position distribution of the Rubidium atoms in the Argon or Neon cold matrix. Finally, we consider theoretically the case of Lithium atoms in a parahydrogen cold matrix, and estimate the relevant interaction energies, and the expected inhomogeneous spectral broadening in such a case. This theoretical analysis leads us to suggest that the combination of dilute Lithium atoms in a parahydrogen cold matrix could be a promising candidate for minimizing the inhomogeneous line broadening resulting from the guest-atoms-matrix interaction in the proposed axion detection scheme. Finally, section \ref{sec4} is devoted to our conclusive remarks.

\section{\label{sec:experimental}Dilute Rubidium atoms in a solid Argon and Neon matrix: experimental results}

In this section we report about the growth and spectral measurements of dilute Rubidium atoms in Argon and Neon solid matrices at cryogenic temperatures.
Dependence of the spectra of solid Argon doped with Rubidium atoms from crystal-growth temperature and annealing history, has been reported in \cite{Kanagin-Regmi13}.

\begin{figure}[!htbp]
\centering\includegraphics[width=8.6cm]{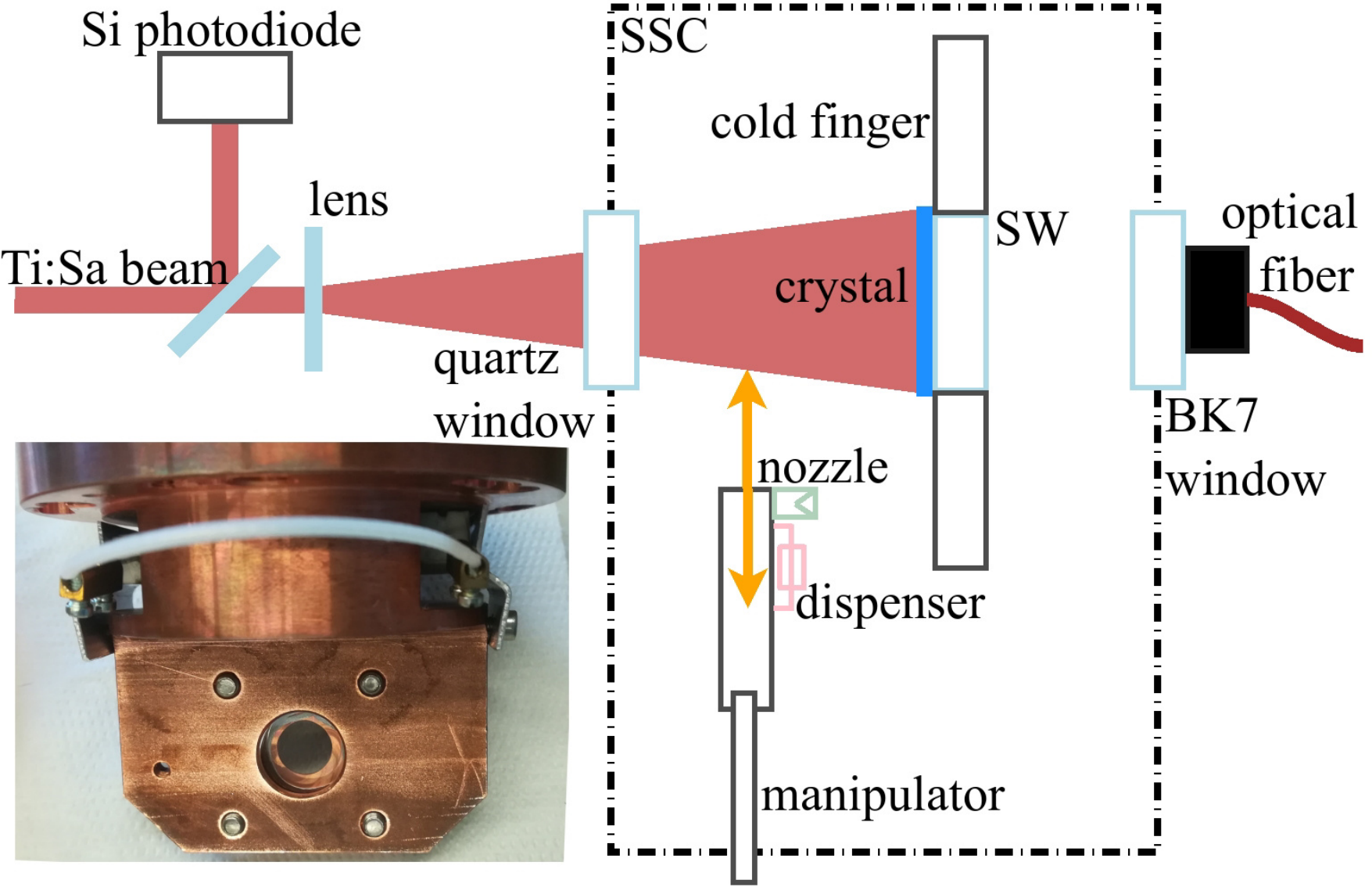}
\caption{Sketch of experimental setup. The inset shows a picture of the cold finger with the sapphire window placed at its centre.}
\label{fig:set-up}
\end{figure}

The core of the experimental apparatus is the cold finger of a pulse tube refrigerator\,\footnote{Sumitomo RP062B} that could reach a minimum temperature of 4\,K allowing thus to grow and to maintain rare gas crystals. The cold finger is placed in the middle of a six-holes cross stainless-steel chamber (SSC) and at its center is mounted a 25\,mm diameter sapphire window (SW) that holds the crystal and allows optical transmission measurements. The SW temperature is monitored through a silicon diode\,\footnote{Lakeshore DT670}, and its precise control is ensured using a 15\,W heater placed nearby the SW.
A quartz and a BK7 windows are placed in the two ports of the SSC along the normal direction to the surface of the SW, while the port in the transverse axis is equipped with a mechanical manipulator where both a gas nozzle and an alkali metal dispenser are installed.
The manipulator can be placed in front of the SW or 10\,cm backwards depending if the growth or the measurements are ongoing. A sketch and a picture of the apparatus described above are shown in figure\,\ref{fig:set-up}.
Both vacuum and pressure of gas into the SSC are measured using pressure gauges in the range $[10^{-9}-100]$\,mbar\,\footnote{Pfeiffer PKR251}.

Crystal growth occurs through spraying the purified rare gas mixed with alkali atoms onto the cold surface of the SW.
Prior to be sprayed, the gas passed within a purification system that allows an impurity contamination below the ppb level especially for high-electronegativity elements. This system is described with more specific details in a previous paper\,\cite{Guarise-Braggio17}.
The Rubidium dispenser\,\footnote{SAES Getter} is connected through two vacuum-feedthroughs to a power supply that delivers up to 6\,A necessary to sublimate the metal. The parameters that have been used to control crystal formation are the partial pressure of the gas and of the dopant during the growing (P$^{gas}_g$, P$^{dop}_g$), the SW temperature (T$_g$) and the time of growing (t$_g$).

Initially, we grew a $\sim100\,\mu$m slab of pure rare-gas with a flow rate maintained at about 1\,mbar$\cdot$l/s that leads in P$^{gas}_g\sim5\cdot10^{-5}$\,mbar.
Only after $\sim30$\,min the doping dispenser has been activated with a current set to $\sim3.5$\,A which gives P$^{dop}_g\sim5\cdot10^{-7}$\,mbar. In such a way, Rubidium atoms have been embedded into Argon and Neon matrices respectively with a ratio between the two species of about $1\%$.
Finally, also in the last $\sim100\,\mu$m we deposited only RG, indeed during this step the dispenser current has been set to zero.
Given the total time of growth $t_g\sim4$\,h, the thickness of the Rb-Ar and Rb-Ne films are about 1\,mm.

During the growth, T$_g$ has been set to $\sim60\,$K and $\sim10$\,K for Argon and Neon respectively, while after crystals formation, an annealing process of $\sim1$\,h at a temperature T$_g+10$\,K has been performed. Finally, for the measurements,  the temperature was lowered to about 5\,K for all the crystals.
Three different samples have been grown to check the reproducibility of any matrix.

The optical part of the experimental set-up is mainly composed of a Titanium-sapphire (Ti:Sa) CW laser tunable in the near-infrared (NIR) region, whose output power varies between 150\,mW and 800\,mW depending on the wavelength. This laser system has been used to cover the broadest range for crystal excitation.
Appropriate quartz lenses have been adopted to couple the light coming from the laser source to the crystal and a small fraction ($\lesssim$\,1$\%$) of the beam has been focused into a silicon photodiode that acts as a laser system monitor. No control of the polarization of the source has been done.
An optical fiber connects the output BK7 window to the the light-detection system that can be made of a NIR spectrometer\,\footnote{Ocean Optics NirQuest} or of a Silicon (Si) photodiode coupled with long-pass filters.
Through these kind of sensors we can measure the fluorescence spectra in different visible and NIR regions.
To minimize the background noise, the measurements have been carried out in a dark environment.
We report here only the measurements relevant for the analysis of the width of the emitted spectrum, and for the comparison with the theoretical calculations that will be presented in the following section.

\begin{figure}[!htbp]
\centering\includegraphics[width=9.3 cm]{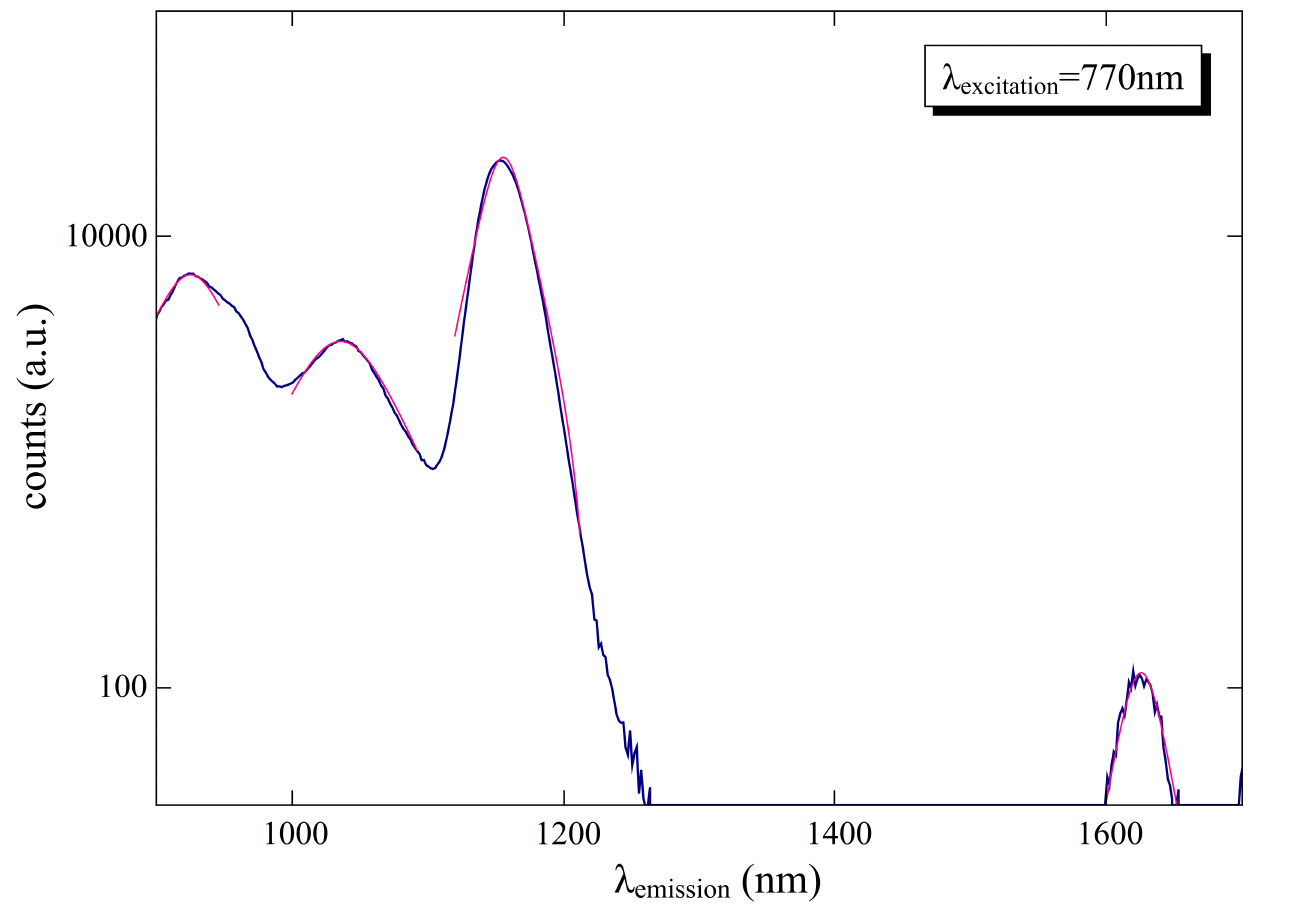}
\caption{Fluorescence spectrum of the Rb-Ar crystal excited with 770\,nm photons. The blue line represents experimental data, while purple curves are the Lorentzian fits.}
\label{fig:ArRb770}
\end{figure}

Figure\,\ref{fig:ArRb770} shows the fluorescence spectrum in the NIR region of the Rb-Ar crystal excited at $\lambda_{excitation}=770$\,nm.
At this excitation wavelength, the emission spectrum is optimized in amplitude; furthermore this $\lambda$ is very close to the D$_{2}$ line of the rubidium \cite{moi2012}.
The blue line represents the acquired data while purple curves are the Lorentzian fits of the peaks, given by:
\begin{equation}
\label{Lorentzian}
L(\lambda ) = y_0 + \frac{2A \Delta \lambda/\pi}{4(\lambda -\lambda_c)^2+ (\Delta \lambda )^2},
\end{equation}
where $\lambda_c$, $A$, $\Delta \lambda$ are respectively the central wavelength, the width and the area of the Lorentzian, and $y_0$ is an offset. The results of the fits are listed in the table\,\ref{tab:exc770}.

\begin{table}[h]
\caption{\label{tab:exc770}Peak parameters of the fits of fig.\,\ref{fig:ArRb770} }
\begin{tabular}{l|l|l|l}
\hline
  \textbf{Peak} &\textbf{ \emph{A\,(a.u.)}} &\textbf{\emph{$\lambda_c$\,(nm)}} & \textbf{ \emph{$\Delta \lambda$\,(nm)}}  \\
  \hline
  1 &(70$\pm$3)$\cdot10^{5}$ &   927.4$\pm$0.3   &71$\pm$10  \\
 2 &(56$\pm$5)$\cdot10^{5}$&    1034.9$\pm$0.3    &92$\pm$4   \\
 3 &(151$\pm$3)$\cdot10^{5}$ &  1154.8$\pm$0.2   &39$\pm$10   \\
 4 & (79$\pm$4)$\cdot10^{2}$&  1624.7$\pm$0.4  &37$\pm$2  \\
   \hline
 \end{tabular}
\end{table}

If we consider the three main peaks of the emission spectrum (peak 1, 2 and 3 of Table \ref{tab:exc770}), their relative energy widths $\Delta E /E = \lvert \Delta \lambda /\lambda_c \rvert$ are:
\begin{eqnarray}
\label{width}
\left( \Delta E /E \right)_1 &\sim& 7 \% \, ,
\left( \Delta E /E \right)_2 \sim 9 \% \, \, \, \nonumber \\
\left( \Delta E /E \right)_3 &\sim& 4 \%
\end{eqnarray}
We can identify these peaks as energy-shifted transitions from the ground state of the Rubidium atoms to the first excited-states group \cite{BASD}, with energy levels corrected by the energy shifts due to the matrix interaction (the unperturbed transition wavelengths from ground to the first upper excited states are in the range $\lambda \sim\,[770 - 795]$\,nm, according to the specific fine and hyperfine transition).
This is very reasonable, because the transitions of the Argon atoms in the solid matrix are at a quite higher energy.

We can meaningfully identify the large broadening of these peaks of the spectrum as the effect of an inhomogeneous broadening of the energy levels, for both the ground and the first electronically-excited levels, consequent (in part, at least) to the interaction of the alkali atoms with the Argon atoms of the solid matrix.

\begin{figure}[!htbp]
\centering\includegraphics[width=9.3cm]{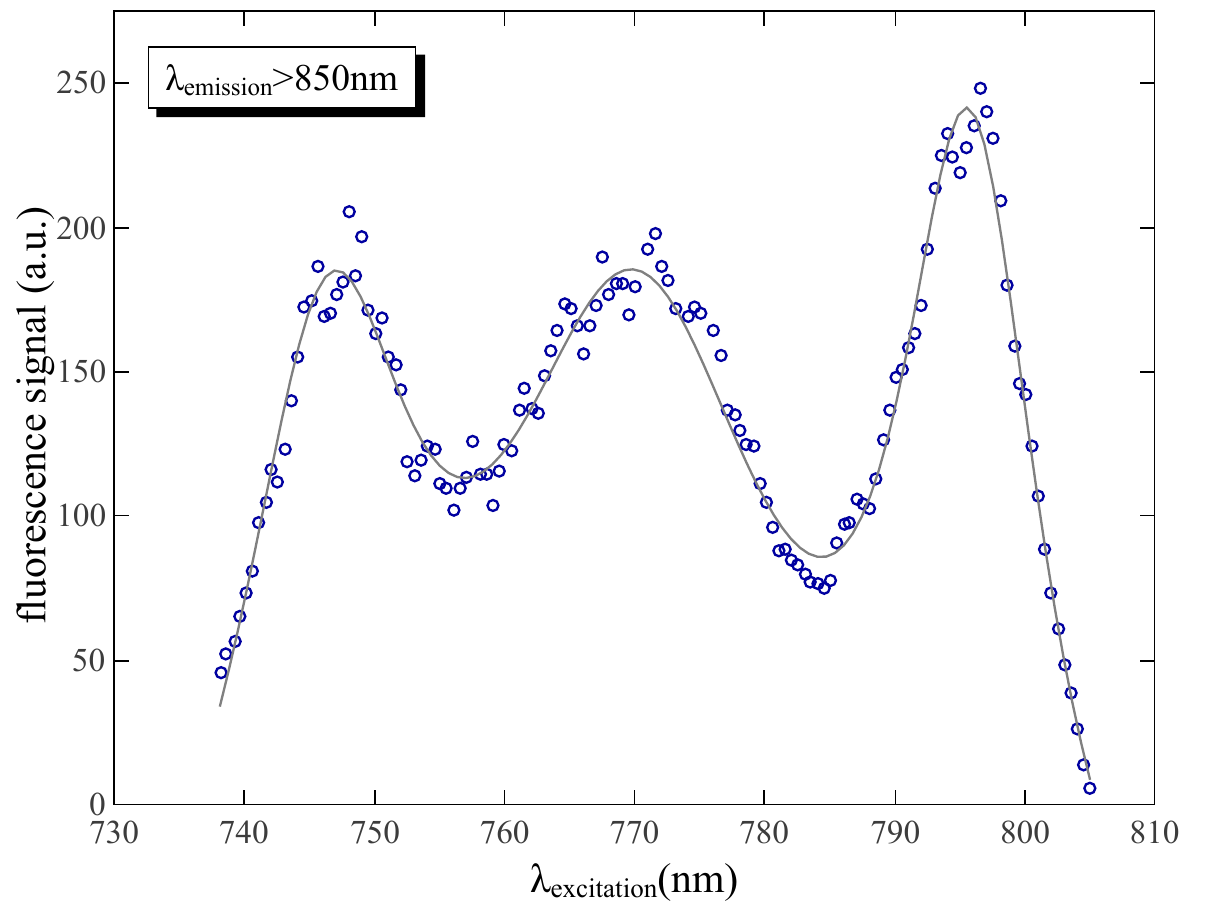}
\caption{Spectrum of Rb-Ar obtained integrating the fluorescence for $\lambda_{emission}>$850\,nm, and changing the excitation of the laser pump in the NIR range. Blue dots are experimental data, while the gray line is the three Lorentzian fits obtained.}
\label{Fig:ArRb850}
\end{figure}

Figure\,\ref{Fig:ArRb850} shows the measured fluorescence emission integrated for $\lambda_{emission}>$850\,nm as a function of the excitation wavelength. Also in this case the experimental data have been fitted with Lorentzian curves and the corresponding parameters are listed in the table\, \ref{tab2}. In this case the relative widths of the three peaks are relatively smaller with respect of the previous case ($ \sim 3 \%$ for the two peaks at lower wavelength, and $\sim 2 \%$ for the third peak), allowing the same physical considerations given above.

\begin{table}[h!]
\centering\caption{Peak parameters obtained from the fits of figure\,\ref{Fig:ArRb850} ($\lambda_{emission}>$850\,nm) }
\begin{tabular}{l|l|l|l}
\hline
  \textbf{Peak} &\textbf{ \emph{A\,(a.u.)}} &\textbf{\emph{$\lambda_c$\,(nm)}} & \textbf{ \emph{$\Delta \lambda$\,(nm)}}  \\
  \hline
  1 &10028$\pm$60 &   748.01$\pm$0.01   &21.34$\pm$0.08  \\
 2 &14288$\pm$70&    772.63$\pm$0.02    &28.05$\pm$0.09   \\
 3 &86233$\pm$50 &  798.022$\pm$0.008   &15.28$\pm$0.06   \\
   \hline
 \end{tabular}
 \label{tab2}
\end{table}

\begin{figure}[!htbp]
\centering\includegraphics[width=9.3cm]{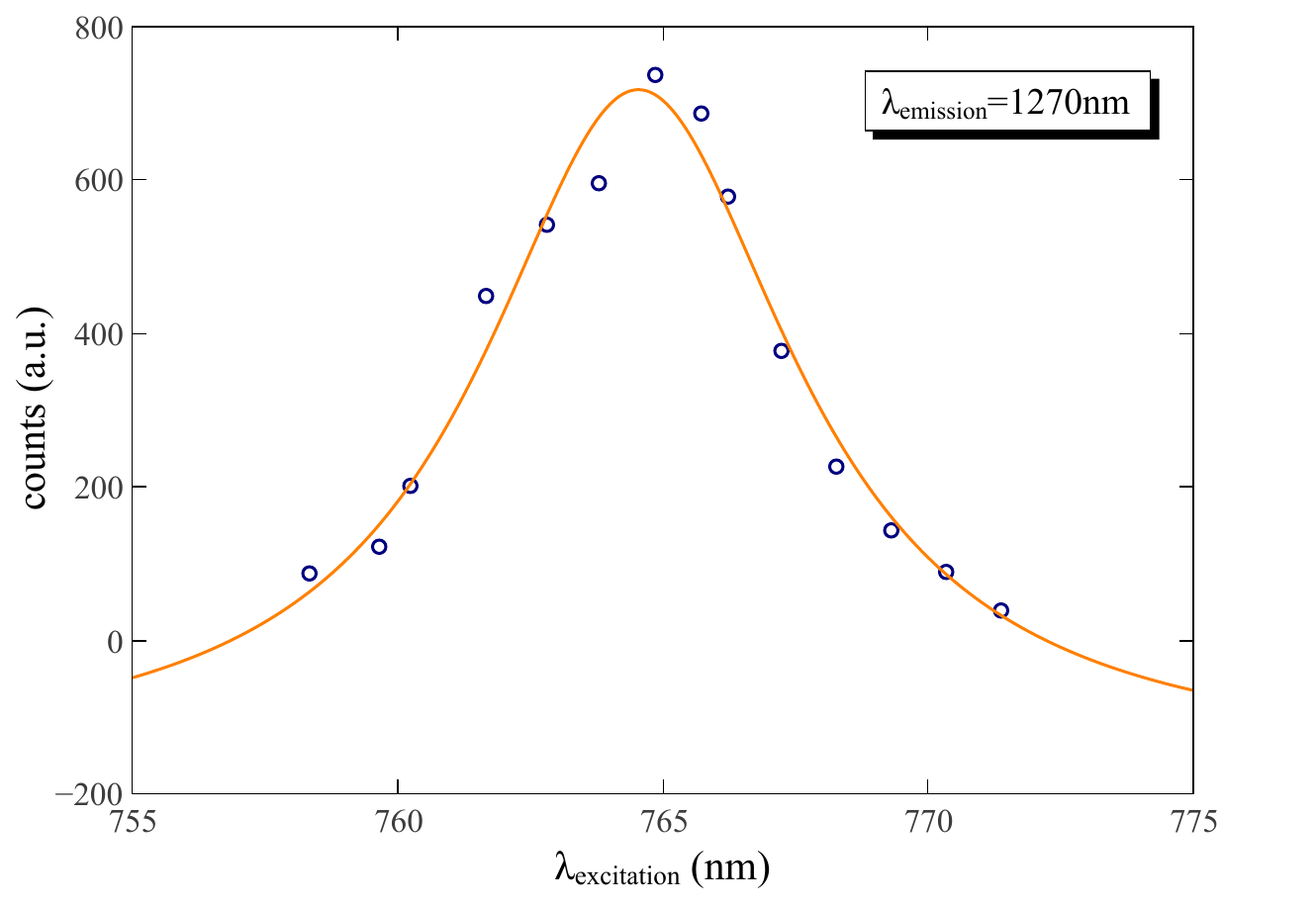}
\caption{Blue dots are the spectrum of the Rb-Ne crystal, observing the light output at $\lambda_{emission}$=1270\,nm as a function of the excitation wavelength. The orange curve is the Lorentzian $L(\lambda)$ fit of the data.}
\label{Fig:RbNe}
\end{figure}

Figures and \ref{fig:ArRb770} and \ref{Fig:ArRb850} show the typical triplet structure of the spectra of atoms embedded in solid matrices, where one peak is usually attributed to the free transition between the \textsuperscript{2}S and the \textsuperscript{2}P levels of the guest atoms, plus two blue-shifted transitions of some tens nanometers \cite{Fajardo-Carrick91}, usually attributed to the splitting of the excited P levels of the guest atom due to the interaction with the matrix. The common interpretation of this triplet structure is that the three peaks refer to guest atoms in the same trapping site, while eventual other triplets in the spectrum are relative to other trapping sites \cite{Fajardo-Carrick91}. This gives support to our working hypothesis in the theoretical analysis in the next section, where we assume a specific trapping site of the doping atoms in the cold inert-gases matrix.

We also report some measurements of the spectra of Rubidium atoms embedded into the Neon matrix.
Figure \ref{Fig:RbNe} shows the spectrum of the Rb-Ne crystal, observing the emitted light at $\lambda_{emission}$=1270\,nm as a function of the excitation wavelength from 755\,nm to 775\,nm. The relative width of the line, which is about 1 \% (smaller than in the Argon case), should imply that the interaction energy of the Rubidium atoms with Ne matrix is smaller, or that the inhomogeneity of the atoms distribution in this crystal is smaller than in Argon.
Possible theoretical interpretations of these results in terms of interaction energies between the guest atoms and the solid matrix will be discussed in the next section.

\section{Dilute alkali atoms in a solid inert-gas matrix: theoretical analysis}
\label{sec2}

\subsection{The theoretical model}
\label{subsec:2.1}

We now consider some effects of a solid matrix of inert atoms or molecules at cryogenic temperatures on embedded dilute alkali atoms.
In this section we estimate the potential energy of a guest alkali atom in the solid matrix, starting from tabulated values of the relevant parameters, obtained experimentally or theoretically calculated.
Our aim here is to obtain estimates of the order of magnitude of the potential, rather than precise values for specific cases, using simple expressions of the dependence of the potentials from the distance, depending from a small number of parameters, and that can be easily adapted to different atom-matrix combinations.
We base our analysis on three different analytical models for the potential energy curve of the doping alkali atom and one atom or molecule of the matrix; specifically, we will consider Argon, Neon and parahydrogen as matrices. The models considered and compared are the Morse potential \cite{Jensen17,Morse29}, the Buckingham-Hill potential \cite{Jensen17,Buckingham38,Hill48}, and the model used by Buck et al for
H\textsubscript{2}-H\textsubscript{2} or D\textsubscript{2}-H\textsubscript{2} interactions \cite{Buck-Huisken83}, and then used also for the Li-H\textsubscript{2} potential \cite{Cheng-Whaley96,Hobza-Schleyer84}).
Our analysis will refer to only a specific peak of the triplet structure of the spectrum, in particular the non-shifted one as described in the previous section.

Evaluating these potentials through simple analytical expressions is an important first step for estimating the contribution to the inhomogeneous linewidth of the energy levels of the dilute guest atoms in the matrix  due to their interaction energy with the matrix. In fact, the actual potential energy depends on the position of the atoms in the solid crystal, which can be (slightly) different from atom to atom, yielding different energy-level shifts; this eventually yields a relevant contribution to the resulting inhomogeneous width of the spectral lines.

In the case of our interest, the Buckingham-Hill and the Morse potentials can give a reliable prediction only in different distance ranges: large distances for the Buckingham-Hill potential, and short distances for the Morse potential.
More sophisticated models such as the one proposed by Buck et al., have been used in the literature for the Li-H\textsubscript{2} potential as well as for the H\textsubscript{2}-H\textsubscript{2} potential \cite{Buck-Huisken83,Cheng-Whaley96}, but they involve a higher number of parameters; we will consider this model for the Li-H\textsubscript{2} potential at the end of this subsection.

As we will now discuss in detail, both of them can be in general necessary in order to evaluate the potential felt by the alkali atom in the solid matrix using a small number of known and tabulated values of the relevant parameters for the potential. It can be also helpful to use an appropriate combination of both models in order to obtain an approximated analytical expression of the actual potential, assuming that the alkali atom substitutes one of the atoms or molecules of the solid matrix. This potential, together with our experimental spectroscopical data of section \ref{sec:experimental}, will be then exploited in subsection \ref{subsec:2.2} to discuss the consequent
contribution to the inhomogeneous broadening of the alkali-atom spectral lines (in particular, of the ground state, involved as initial state of the proposed transition scheme for the axion detection discussed in the Introduction.

We first consider the case of the doped crystal experimentally analyzed in the previous section, that is dilute Rubidium atoms in a Argon or Neon solid matrix, comparing the theoretical results with our experimental data reported in section \ref{sec:experimental}. Secondly, we will consider the case of dilute Lithium atoms in a parahydrogen (p-H\textsubscript{2}) solid matrix, obtaining insights that this alkali-matrix system yields quite smaller interaction energies, at least of the ground state. We finally suggest that this could be a good guest-atoms-matrix combination for the proposed axion detection scheme.

Both the inert-gas atoms of the cold matrix and the doping alkali atoms
have a spherical electronic structure; thus, outside the wavefunction overlap region, the only potential is the van der Waals potential, with a leading term scaling with the distance as $r^{-6}$ (dipolar two-body dispersion interaction energy) \cite{Craig-Thirunamachandran98,Salam10,Passante18}. We can thus write the total two-body potential between a ground-state alkali atom (spherically symmetric) and one atoms of the solid matrix in the following form (the same is valid in the case of Li-(p-H\textsubscript{2}) we will consider successively)
\begin{equation}
V(r)= V_{vdw}(r) + V_{rep}(r) ,
\label{eq:potential}
\end{equation}
where $V_{vdw}(r)$ is the attractive two-body van der Waals potential, and $V_{rep}(r)$ is a repulsive exchange or Coulomb potential, acting at short distances when a significant wavefunction's overlap occurs. Several forms of these potentials have been proposed and used in the literature. We now introduce some relevant features of three of them, that we will use in the theoretical analysis given in the next subsection.

Taking the usual $r^{-6}$ dipolar attractive van der Waals potential energy for ground state systems, and a simple exponential for the repulsive short-distance component, we have the Buckingham-Hill potential \cite{Jensen17,Buckingham38,Hill48}
\begin{equation}
V_{\text{BH}}(r)= -\frac {C_6}{r^6} + A e^{-r/{R_{rep}}} ,
\label{eq:Buckingham-Hill}
\end{equation}
where the three parameters $C_6,A,R_{rep}$, i.e. van der Waals constant, strength and distance scale of the repulsive potential, respectively, appear.

We assume that the total potential felt by the alkali atom inside the solid matrix is just the sum of the two-body potentials due to the first neighbors molecules of the matrix, thus neglecting many-body effects.
In the Buckingham-Hill potential, we also neglect higher multipolar dispersion (long-range) interactions \cite{Salam10}, scaling with the distance as $r^{-8}$ and higher inverse powers of the distance \cite{Patil91,Mitroy-Zhang07}. These approximations are justified by the relatively large reticular distance,
$a_{Ar} = 5.26 \, \text{\AA}$ for Ar and $a_{Ne} = 4.43 \, \text{\AA}$ for Ne \cite{Guarise-Braggio20,Xu-Hu11,Dobbs-Jones57,Ryan-Collier10}.

For our purposes, it is important to relate the constants in the Buckingham-Hill potential (\ref{eq:Buckingham-Hill}) to tabulated experimentally measured or theoretically computed quantities, such as the dimer equilibrium distance and dissociation energy (that is the potential energy at the equilibrium position). The van der Waals constant $C_6$ is tabulated for many atoms and molecules in their ground or in excited states \cite{Mitroy-Zhang07,Zhu-Dalgarno02}.

In order to evaluate the constant $A$ in (\ref{eq:Buckingham-Hill}) in terms of known parameters, we can impose that
$\partial V_{\text{BH}}(r)/\partial r =0$ for $r=R_{eq}$, $R_{eq}$ being the equilibrium distance of the alkali-inert-atom dimer. We immediately find
\begin{equation}
\label{constantA}
A=\frac {6C_6R_{rep}}{R_{eq}^7}e^{R_{eq}/R_{rep}} ,
\end{equation}
yielding
\begin{equation}
\label{Hill}
V_{\text{BH}}(r)= \frac {6C_6R_{rep}}{R_{eq}^7}e^{(R_{eq}-r)/R_{rep}} - \frac {C_6}{r^{6}} .
\end{equation}

By taking into account that the quantities experimentally measured are the van der Waals constant $C_6$, the equilibrium distance $R_{eq}$, and the dissociation energy $V(R_{eq})$ (that is the potential energy at the equilibrium distance), from (\ref{Hill}) we can obtain  the following expression for $R_{rep}$
\begin{equation}
\label{Rrep}
R_{rep} = \frac {R_{eq}^7}{6C_6} \left( V_{\text{BH}}(R_{eq}) + \frac {C_6}{R_{eq}^6} \right) .
\end{equation}

The complete expression of the potential is then
\begin{equation}
\label{eq:Buckingham-Hill1}
V_{\text{BH}}(r) = - \frac {C_6}{r^6} + \left( V_{\text{BH}}(R_{eq}) + \frac {C_6}{R_{eq}^6}\right)
\exp \left[ \frac {6C_6}{R_{eq}^7 (V_{\text{BH}}(R_{eq})+ \frac {C_6}{R_{eq}^6})} (R_{eq} -r)\right]
\end{equation}

Thus, using Eq. (\ref{eq:Buckingham-Hill1}), the Buckingham-Hill potential is entirely expressed in terms of hopefully known and tabulated parameters (for example, the various parameters for several alkali and RG atoms can be found in \cite{Mitroy-Zhang07,Rossi-Pascale85,Zhu-Dalgarno02}).

The Buckingham-Hill potential is a good approximation for distances around $R_{eq}$, yielding a good approximation to the equilibrium distance, and for larger distances, where it correctly reduces to the van der Waals potential, but not for shorter distances: in fact, for distances smaller than $R_{eq}$ the Buckingham-Hill potential decreases to negative values, while it should grow and give the short-distance repulsion. In other words, the equilibrium position of the potential (\ref{Hill}) is an unstable point.

We thus need an appropriate approximation to the potential energy valid for distances smaller than the equilibrium distance. As it will emerge in the next subsection, the short-distance case is that we are mainly interested to.
In the short distance region, that is for distances smaller than the dimer's equilibrium distance, we can use the Morse potential, given by \cite{Jensen17,Morse29}

\begin{equation}
\label{Morse}
V_{\text{M}}(r)= D \left( 1- e^{-\alpha(r-R_{eq})} \right)^2 -D ,
\end{equation}
$D=\lvert V_{\text{M}}(R_{eq}) \rvert$ being the dissociation energy of the dimer, that is the energy at the equilibrium distance $R_{eq}$, $\alpha=\sqrt{\kappa /(2D)}$, with $\kappa =\partial^2V(r)/\partial r^2 \lvert_{r=R_{eq}}$ the curvature of the potential at the equilibrium distance; $\kappa$ is a second parameter characterizing the interatomic potential energy.

Although the Morse potential (\ref{Morse}) is a good approximation to the short-range exchange/Coulomb potential, it fails at large distances.
The Buckingham-Hill and the Morse potential are both expected to give a reasonably good approximation around the equilibrium distance.

The Morse potential (\ref{Morse}) contains three free parameters, but its mathematical expression does not allow to obtain a relation between them exploiting that its derivative must vanish at the equilibrium distance $R_{eq}$.
For this reason, since the constant $\kappa$ (that is the curvature of the potential at the equilibrium distance) is not always available from tabulated data in the literature,
we suggest that in such cases we can resort to the Buckingham-Hill potential for its estimation. This could be done by calculating the modulus of the second derivative of the Buckingham-Hill potential at the equilibrium distance, and finally obtain the parameter $\alpha$ in the Morse potential (\ref{Morse}). We stress that some care should be used in this procedure, because, at the equilibrium distance, the BH potential has a maximum while the Morse potential has a minimum. Thus, using in the Morse potential the value of $\kappa$ obtained from the BH potential, assumes that, around the equilibrium distance, the modulus of the curvature of both potentials are close to each other.
Assuming this, from (\ref{eq:Buckingham-Hill1}) we get
\begin{equation}
\label{curvature}
\kappa = \left| \frac {\partial V_{\text{BH}}(r)}{\partial r^2} \rvert_{r=R_{eq}} \right|
= \frac {6C_6}{R_{eq}^8} \left| \frac {6}{\frac{R_{eq}^6V_{\text{BH}}(R_{eq})}{C_6}+1} -7 \right| .
\end{equation}

Because in our case the distance between the alkali atom and any inert-gas atom molecule of the solid matrix is smaller than the equilibrium distance ($3.72 \, \text{\AA}$, to be compared to $5.498 \, \text{\AA}$, for the Rb-Ar dimer, for example),
we expect that the Morse potential gives a better approximation to the true potential than the Buckingham-Hill potential. How good is this procedure for obtaining the parameter $\kappa$ can be checked in the cases where $\kappa$ is known, experimentally or theoretically, from independent considerations, as we will explicitly discuss later on for Rubidium atoms in a solid Argon or Neon matrix.

A third potential, accurate for the Lithium-parahydrogen dimer and that we will use in the next subsection for this case, although it involves a larger number of parameters, has been discussed and used in \cite{Buck-Huisken83,Cheng-Whaley96}. This potential has the following expression
\begin{equation}
\label{Buck}
V_B^{Li-H\textsubscript{2}}(r) = e^{\alpha -\beta r -\gamma r^2} -f(r) \left( \frac{C_6}{r^6} + \frac{C_8}{r^8} + \frac{C_{10}}{r^{10}} \right),
\end{equation}
with
\begin{equation}
\label{funct}
f(r) = \left\{
\begin{array}{cc}
e^{-(1.28r_m/r-1)^2} & \mbox{for $r < 1.28 \, r_m$} \\
1 & \mbox{for $r > 1.28 \,r_m$ .}
\end{array}
\right. ,
\end{equation}
where $\alpha$, $\beta$, $\gamma$, $r_m$ are parameters relative to the atomic and molecular species considered, while $C_6$, $C_8$ and $C_{10}$ are the relative van der Waals constants. This potential has a spherical symmetry, because non-spherical components, depending from the specific geometrical configuration of the Li-H\textsubscript{2} system, are small; the potential (\ref{Buck}) is accurate in the distance range $r \sim 2-12$ \AA . It has been also used for the H\textsubscript{2}-H\textsubscript{2} and H\textsubscript{2}-D\textsubscript{2} interaction energies \cite{Buck-Huisken83,Cheng-Whaley96}.

\subsection{Numerical estimates of the atoms-matrix interaction energies}
\label{subsec:2.2}

We can now use the potentials of subsection \ref{subsec:2.1} to obtain approximated estimates of the
the value of the interaction energy between an alkali atom (Lithium, Sodium, Potassium, Rubidium) and an atom or molecule of the solid matrix (in this section we will consider Rubidium in Argon and Neon, and then Lithium in parahydrogen), assuming that the alkali atom replaces one atom or molecule of the solid crystal.
From that, the interaction energy between a guest alkali atom and the solid RG or parahydrogen matrix can be easily obtained, assuming only pairwise interactions between the nearest neighbours. As mentioned before, this should be a fairly good approximation due to the large reticular distance of the solid matrix.

Possible trapping sites of alkali atoms in RG crystals, and their stability, have been recently studied in the literature \cite{Ahokas-Kiljunen00,Ozerov-Bezrukov19,Tarakanova-Buchachenko20}.

We first consider the Rb-Ar case and then the Rb-Ne case, comparing our theoretical estimates with the measured spectra reported in the previous section. At the end, we will also analyze the Li-(p-H\textsubscript{2}) case.
Throughout this subsection, the units used are eV for the energy and $\text{\AA}$ for the distance.

Solid Argon and solid Neon have both a face-centered cubic (fcc) crystal structure; the reticular distance is $a_{Ar} = 5.26 \, \text{\AA}$ for Ar and $a_{Ne} = 4.43 \, \text{\AA}$ for Ne \cite{Guarise-Braggio20,Xu-Hu11,Dobbs-Jones57,Ryan-Collier10}. The guest alkali atom considered is Rubidium (Rb), as in the measurements reported in the previous section.
We assume that the Rubidium atom substitutes one of the atoms of the matrix.
Taking into account the fcc reticular structure of Argon and Neon, each Rubidium atom interacts with 12 atoms of the crystal at a distance $a_{Ar(Ne)}/\sqrt{2}$, $a$ being the reticular distance.

In order to evaluate the potential energy through the Morse or the Buckingham-Hill potential, we need the numerical values of the equilibrium distance $R_{eq}$, the dissociation energy $D= \left| V(R_{eq}) \right|$, the curvature $\kappa$ of the potential energy at the equilibrium distance or the van der Waals constant $C_6$.
Theoretical estimates of the equilibrium distance $R_{eq}$ and the dissociation energy $D$ can be found in several works \cite{Baylis69,Patil91,Goll-Werner06,Blank-Weeks12}. The curvature of the potential energy at the equilibrium distance can be found in \cite{Baylis69,Goll-Werner06}.
The $C_6$ coefficients for several alkali atoms and RG atoms, both in the ground state and in excited levels, can be found in \cite{Zanuttini-Jacquet09,Mitroy-Zhang07,Zhu-Dalgarno02,Standard-Certain85}.  Although there is a good agreement in the literature on the values of $R_{eq}$ and $C_6$, the tabulated values of $D$ and $\kappa$ for Rb-Ne can significantly differ in different papers, and even in the same paper when different methods of calculation are used; the available data for Rb-Ar are instead much more uniform. For this reason we will concentrate mainly on the Rb-Ar case. We have chosen to use the data reported in \cite{Goll-Werner06}, where the three necessary parameters $R_{eq}$, $D$, $\kappa$ are calculated. We will also compare these data with those reported in \cite{Blank-Weeks12}. However, the small variability of the tabulated data for Rb-Ar is not much relevant for our purposes, because we only need an order-of-magnitude estimate of the interaction energy, and not its exact value.

\vspace{6pt}

{\it i) Rb-Ar}

\vspace{3pt}

We first consider the ground-state configuration $X^2\Sigma^+_{1/2}$ of the Rb-Ar dimer; it corresponds to the electronic configuration of both the Rubidium and Argon atoms in the ground state. The relevant data are \cite{Goll-Werner06}
\begin{eqnarray}
\label{data-Ar}
R_{eq}^{Rb-Ar}&=& 5.498 \, \text{\AA} , \nonumber \\
D^{Rb-Ar} &=& 4.45 \cdot 10^{-3} \, \text{eV} , \nonumber \\
\kappa^{Rb-Ar} &=& 5.96 \cdot 10^{-3}  \, \text{eV} {\text{\AA}^{-2}} .
\end{eqnarray}
For the same configuration, the data in \cite{Blank-Weeks12} for the equilibrium distance and the dissociation energy (the curvature of the potential is not explicitly given) are not very dissimilar:
$R_{eq}^{Rb-Ar}= 5.482 \, \text{\AA}$, $D^{Rb-Ar} = 6.07 \cdot 10^{-3} \, \text{eV}$.

As mentioned above, we need to evaluate the potential energy at the distance
$a^{Rb-Ar}=a_{Ar}/\sqrt{2}= 3.72 \, \text{\AA}$, assuming that the Rubidium atom replaces one Argon atom of the matrix.
Since $a^{Rb-Ar}<R_{eq}^{Rb-Ar}$, we use the Morse potential (\ref{Morse}). The $\alpha=\sqrt{k/(2D)}$ parameter in Eq. (\ref{Morse}) in the present case is therefore $0.818 \, {\text{\AA}^{-1}}$. Using all these data in (\ref{Morse}), we immediately find
\begin{equation}
\label{potential:Rb-Ar}
V_{\text{M}}^{Rb-Ar} (r=a^{Rb-Ar}=3.72 \, \text{\AA}) \simeq 4.35 \cdot 10^{-2} \, \text{eV} .
\end{equation}

In this case we have been able to estimate the potential energy using the Morse potential only, because the potential curvature $\kappa$ is known. In subsection \ref{subsec:2.1} we proposed that $\kappa$ could be obtained using the Buckingham-Hill potential through Eq. (\ref{curvature}). The van der Waals constant $C_6$ can be found in \cite{Mitroy-Zhang07,Patil91}, for example, as
$C_6^{Rb-Ar}=201 \, \text{eV} {\text{\AA}}^{6}$. From Eq. (\ref{curvature})
we thus obtain $\kappa_M = \kappa^{Rb-Ar} = 4.73 \cdot 10^{-3}  \, \text{eV} {\text{\AA}}^{-2}$, that is relatively close to the tabulated value shown in (\ref{data-Ar}). This is an important consistency check for the analytical method presented in this section.

Summing the potential (\ref{potential:Rb-Ar}) over the 12 nearest-neighbors of the Rubidium atom (pairwise summation), we get
\begin{equation}
\label{total-Rb-Ar}
V_{tot}^{Rb-Ar} \simeq 0.52 \, \text{eV} .
\end{equation}

Using the very recent data for the equilibrium distance and the dissociation energy in \cite{Medvedev-Meshkov18}, $R_{eq}^{Rb-Ar}= 5.541 \, \text{\AA}$, $D^{Rb-Ar} = 4.716 \cdot 10^{-3} \, \text{eV}$,  and the data for the curvature of the potential at the equilibrium distance communicated to us by the authors of that paper \cite{Stolyarov21}, $\kappa^{Rb-Ar} = 5.617 \cdot 10^{-3}  \, \text{eV} {\text{\AA}^{-2}}$, we get $V_{\text{M}}^{Rb-Ar}(r=a^{Rb-Ar}/\sqrt{2}=3.72 \, \text{\AA}) \simeq 3.99 \cdot 10^{-2} \, \text{eV}$, that is close to (\ref{potential:Rb-Ar}).
Interpolation of recent ab-initio numerical data for the potential energy \cite{Stolyarov21} (see also \cite{Medvedev-Meshkov18}), yields a somehow lower value, $\simeq 3.1 \cdot 10^{-2} \, \text{eV}$, in any case not far from our analytical estimate.
We can thus say that, even using numerical data from different sources, we obtain essentially the same approximate estimate of $V_{\text{M}}^{Rb-Ar}$.

Even if we are mainly interested to the ground state, as discussed later on, we can use the same procedure to estimate the potential energy of an excited state of the Rubidium atom in the solid Argon matrix. Specifically, we consider the first excited state in the Rb-Ar molecular configuration $A^2\Pi_{1/2}$ (corresponding to a ${\ }^2P_{1/2}$ configuration of Rubidium, and the Argon in its ground state). We use the following tabulated data:
$R_{eq \, (exc)}^{Rb-Ar}= 3.60 \, \text{\AA}$, $D_{(exc)}^{Rb-Ar} = 3.162 \cdot 10^{-3} \, \text{eV}$ \cite{Blank-Weeks12}; $C_{6 \, (exc)}^{Rb-Ar}=325.7 \, \text{eV} {\text{\AA}}^{6}$ \cite{Mitroy-Zhang07}. Because in this case the distance at which we should evaluate the potential, $r=a^{Rb-Ar}/\sqrt{2}=3.72 \, \text{\AA}$, is very close (slightly larger, to be precise) to the equilibrium distance$R_{eq \, (exc)}^{Rb-Ar}$, we can use the Buckingham-Hill potential (\ref{eq:Buckingham-Hill1}), and we obtain
$V_{BH \, (exc)}^{Rb-Ar} \simeq - 3.07 \cdot 10^{-2} \, \text{eV}$.
Since the interatomic distance now considered is close to the equilibrium distance, we must expect that the Morse potential should yield the same result too. An explicit evaluation by using Eq. (\ref{Morse}), with the potential curvature obtained from (\ref{curvature}), gives indeed practically the same result.
After pairwise sum over the 12 nearest neighbors Argon atoms, we have
$V_{tot \, (exc)}^{Rb-Ar} \simeq - 0.37 \, \text{eV}$. These interaction energies are of the same order of magnitude of the ground-state interaction energies evaluated above.

Our main aim here is the role of these interactions in the linewidth broadening of the guest atoms in the solid matrix. Together with other effects, phonon excitations in the matrix for example, they have also other consequences, of course, for example energy shifts due to the degeneracy lifting of excited atomic levels. We will not consider these effects in the present preliminary study. We will limit our considerations only to the contribution of the interactions of the doping atoms with the matrix for their ground state, splitted in the two Zeeman sublevels by the external magnetic field. With reference to the detection scheme of Fig. \ref{Fig1}, this is reasonable if $c$ represents the edge of the crystal conduction band (the axion transition $a \rightarrow b$ does not involve a change in the atomic wavefunction, and thus it should not perturb the crystal structure). If $c$ in Fig. \ref{Fig1} represents an excited level of the doping atoms, we should also add the linewidth of the excited level and the possible perturbation (phonons, for example) due to the change of the atomic wavefunction in the transition between level $a$ or $b$ and the excited level $c$.
For this reason, a detailed comparison of the experimental spectral widths reported in the previous section with our theoretical analysis of the interaction energies between the ground-state guest atom with the cold matrix, is not immediate; however, the present theoretical analysis of this contribution is significant since it can give useful hints on how to minimize this contribution to the width of the ground state, which is the level from which the axion absorption occurs.

We now assume that the contribution to the inhomogeneous broadening of the lines due to the interactions between the dilute alkali atoms and the RG solid matrix is some small fraction of the energy given in (\ref{total-Rb-Ar}), plus eventually that relative to the excited level in the case of a transition between two discrete levels, that we have seen to be of the same order of magnitude for the Rb-Ar system.
Which value has this fraction is very difficult to evaluate from a theoretical point of view only, because it depends on specific details of the position distribution of the alkali atoms inside the solid matrix, for example the inhomogeneity of their position, imperfections of the crystal structures, etc. Thus, it can significantly depend on how the solid matrix is growth, and, importantly, how the dilute alkali atoms are diffused inside the matrix.
We can assume that the main effect is due to a (slight) inhomogeneous distribution of the position of the guest alkali atoms in the RG solid matrix. Comparing our experimental results for the spectral linewidth, reported in section \ref{sec:experimental}, with our present theoretical prediction of the atom-matrix interaction energy, we now show that we can predict a semi-quantitative estimate of the atoms' position distribution in the matrix. This analysis is also based on the fact, known in the literature and mentioned in the previous section, that each triplet of the spectrum is relative to doping atoms in the same trapping site \cite{Fajardo-Carrick91}.

Quantitatively, a spreading of the alkali positions inside the crystal determines an inhomogeneous contribution to the linewidth of the spectrum, which strongly depends on how steep the potential energy is at their average position, of course. Assuming, as before, that the alkali atoms replace one of the atoms of the matrix, being $r_{Rb-Ar} < R_{eq}^{Rb-Ar}$, we can expand the Morse potential (\ref{Morse}) around $r_{Rb-Ar}$ (taking into account the measured relative energy widths given in (\ref{width}), we expect a relatively sharp distribution of the positions). We indicate with $\Delta r$ a small displacement from the average distance $r_0$, which is equal to $r_{Rb-Ar}$ in the present Rb-Ar case, we can obtain the consequent change of the potential energy as
\begin{equation}
\label{expansion}
\Delta V_M(\Delta r) = V_M(r_0+\Delta r)-V_M(r_0)
\simeq 2 \alpha D e^{\alpha (R_{eq}-2r_0)} \left( e^{\alpha r_0}-e^{\alpha R_{eq}} \right) \Delta r ,
\end{equation}
and thus the relative absolute value of the change of the potential energy is
\begin{equation}
\label{expansion2}
\left| \frac {\Delta V_M(\Delta r)}{V_M(r_0)} \right|
\simeq \left| \frac{2 \alpha \left( e^{\alpha r_0}-e^{\alpha R_{eq}} \right)}{e^{\alpha R_{eq}} -2e^{\alpha r_0}} \Delta r \right| ,
\end{equation}
This relation allows us to estimate the position dispersion $\Delta r$ of the alkali atoms in the solid matrix, once the relative dispersion of the potential energy is known. The latter can be obtained from the spectral experimental results of the previous section, specifically from the relative energy widths (\ref{width}). In the case of Rb-Ar here considered, for the peak 3 of table \ref{tab:exc770} (see also Fig. \ref{fig:ArRb770}) with an energy witdh of $4\%$, and using the numerical data in (\ref{data-Ar}), Eq. (\ref{expansion2}) yields the Argon-atoms position dispersion
\begin{equation}
\label{posdisp}
\left| \Delta r \right| \sim 1.7 \cdot 10^{-2} \text{\AA} \, ,
\end{equation}
that means a relative dispersion of $\sim 0.5 \%$. Indeed, this should be considered as an upper limit to the position dispersion around the trapping site, because other effects, not considered here, can contribute to the line broadening.

Our experimental measurements, combined with our theoretical model, have thus allowed us to obtain relevant hints on how the guest atoms in our Rb-Ar experimental setup diffuse in the cold matrix, and relate the measured spectral widths to the alkali-matrix interactions. This method can be also applied also to other doping-atoms-matrix, for example Rubidium atoms in a cold Neon matrix.

\vspace{6pt}

{\it ii) Rb-Ne}

\vspace{3pt}

We now consider the Rb-Ne case, whose spectral measurements have been reported at the end of section \ref{sec:experimental}.
The relevant numerical values for equilibrium distance, dissociation energy and van der Waals constant can be found in  \cite{Baylis69,Patil91,Goll-Werner05,Goll-Werner06,Mitroy-Zhang07,Medvedev-Meshkov18}.
In this case, some differences are found in the literature among the data obtained with different methods, in particular for the dissociation energy. This can yield different estimates of the interaction energy in the solid matrix, depending on the numerical data used.

Solid Neon has a fcc crystal structure as solid Argon, with a reticular distance $a_{Ne} = 4.43 \, \text{\AA}$.
We first use the very recent data of the equilibrium distance and dissociation energy for the Rb-Ne dimer reported in \cite{Medvedev-Meshkov18}, $R_{eq}^{Rb-Ne} = 6.188 \, \text{\AA}$ and $D^{Rb-Ne} = 7.28 \cdot 10^{-4} \, \text{eV}$ respectively, supplemented with the value of the potential curvature at the potential minimum \cite{Stolyarov21}, $\kappa^{Rb-Ne} = 1.107 \cdot 10^{-3} \, \text{eV} {\text{\AA}^{-2}}$. In this case we obtain
\begin{equation}
\label{potential:Rb-Ne}
V_{\text{M}}^{Rb-Ne}(r=3.13 \, \text{\AA}) \simeq 0.13 \, \text{eV} ,
\end{equation}
which is larger than that for Rb-Ar.
A similar result is obtained using the data in \cite{Goll-Werner06}, $R_{eq}^{Rb-Ne} = 6.212 \, \text{\AA}$, $D^{Rb-Ne} = 6.07 \cdot 10^{-4} \, \text{eV}$, $\kappa = 1.061 \cdot 10^{-3} \, \text{eV} {\text{\AA}^{-2}}$, where the value of the curvature $\kappa$ of the potential at the equilibrium position has been inferred from their data of the harmonic wavenumbers, obtaining in this case $V_{\text{M}}^{Rb-Ne}(r=3.13 \, \text{\AA}) \simeq 0.17 \, \text{eV}$.
This is also confirmed by other recent numerical data \cite{Stolyarov21} (see also \cite{Medvedev-Meshkov18}), yielding $V(r=3.13 \, \text{\AA}) = 0.061 \, \text{eV}$: this value is quite lower than our theoretical estimate based on the Morse potential (it is however larger compared to the Rb-Ar case), showing again a larger variability of the energies in the Rb-Ne case according to the analytical or numerical method used.

Although our theoretically predicted values for the ground-state interaction energy for Rb-Ne are larger than for Rb-Ar, as a comparison with (\ref{potential:Rb-Ar}) immediately shows, while our spectra of section \ref{sec:experimental} show sharper linewidths in the Rb-Ne case, this is not necessarily a contradiction between experimental results and theoretical analysis. In fact, also other aspects should be taken into account for a direct comparison of the linewidths. Firstly, our theoretical estimates involve only the ground state, while the spectral linewidths of section \ref{sec:experimental} include also the effects of the excited level, and consequent rearrangement of the wavefunction distribution and possible phonon emission. A wavefunction rearrangement is present also in the considered case of transition to the crystal conduction band, but we expect its effect to be much smaller with respect to the transition to a higher excited level, that involves a larger size of the final discrete state and thus a larger perturbation of the crystal. Thus, we guess that our ground-state linewidth estimates could be directly compared to future measurements of transitions of the doping atom's electron from the ground state (also if Zeeman shifted) to the crystal conduction band.
Secondly, we cannot exclude that the position distribution of the guest atoms in the solid matrix is smaller in Neon compared to Argon.
Indeed, the two solid matrices have different physical properties, related to the interactions between the matrix atoms, for example the condensation temperature, $T_{cond} = 82 \, \text{K}$ for Ar and $T_{cond} = 20 \, \text{K}$ for Ne, and the reticular distance: thus, nontrivial differences in the way the guest atoms arrange themselves in the matrix are not unexpected. We plan to discuss these points in a future work.
In any case, we believe that the case of Neon deserves a more precise analysis from the theoretical side, also because the alkali-RG atoms distance in the Neon matrix is smaller than in the Argon matrix,
in particular in comparison with the corresponding equilibrium distance ($R_{eq}^{Rb-Ar}= 5.498 \, \text{\AA}$, $r=3.72 \, \text{\AA}$ for Rb-Ar; $R_{eq}^{Rb-Ne}= 6.19 \, \text{\AA}$, $r=3.13 \, \text{\AA}$ for Rb-Ne). At shorter distances, the potential energy is very steep,
as, consistently, the plots of the potential energies in \cite{Medvedev-Meshkov18} clearly show. All this makes the details of the model more critical: even a small variation of the relevant parameters can yield a consistent variation of the estimated potential energy.

\vspace{6pt}

{\it iii) Li-(p-H\textsubscript{2})}

\vspace{3pt}

Finally, we consider theoretically the Li-(p-H\textsubscript{2}) case, that is dilute Lithium atoms diffused in a solid parahydrogen matrix. We will obtain a strong indication that this system shows a quite smaller guest-atom-matrix interaction energy.

\begin{figure}[!htbp]
\centering\includegraphics[width=9 cm]{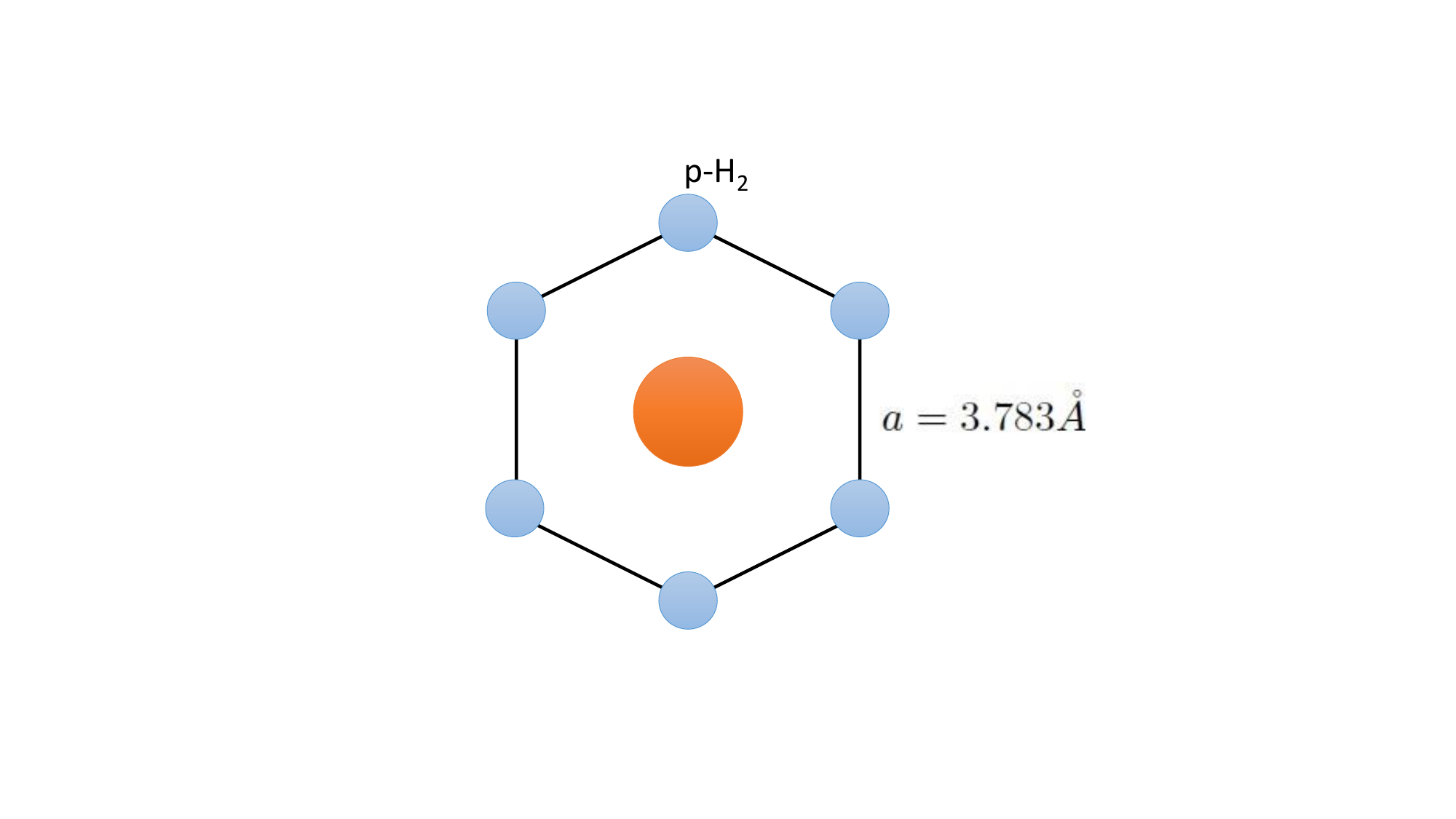}
\caption{The alkali atom at the center of the hexagonal structure of the solid p - H\textsubscript{2} matrix. The alkali atom (orange circle) is assumed to substitute one of the molecules of the solid matrix of p - H\textsubscript{2} molecules (blue circles).}
\label{Fig2}
\end{figure}

The p-H\textsubscript{2} molecule has nuclear spin $I=0$ and, at low temperatures, an even total angular momentum $J$, with a spherical symmetry yielding vanishing permanent multipolar momenta \cite{Bransden-Joachain03}.
Due to the spherical symmetry of the parahydrogen molecule, considerations on the potentials similar to the rare gases, discussed above, can be applied. Solid parahydrogen has a Hexagonal closed packed  (Hcp) structure, with a quite large intermolecular distance compared to the relevant atomic or molecular dimensions. We assume that the alkali atom is placed at the hexagon center, substituting one of the H\textsubscript{2} molecules, as shown in Fig. \ref{Fig2}; due to the large intermolecular distance of the molecules in the p - H\textsubscript{2} matrix, its presence should not change significantly the structure of the crystal. Also, the known large compressibility of parahydrogen solids allows to easily accomodate the alkali-atom impurity \cite{Scharf-Martyna93}. The reticular distance for the parahydrogen crystal is $a \simeq 3.78 \, \text{\AA}$ \cite{Silvera80,Momose-Shida98}.

The relevant tabulated data for the dissociation energy $D$ and the equilibrium distance $R_{eq}$ can be found in \cite{Rossi-Pascale85}, while the van der Waals constant $C_6$ can be found in \cite{Zhu-Dalgarno02}; the tabulated values are $D=3.90 \cdot 10^{-3} \, \text{eV}$, $R_{eq}= 4.76 \, \text{\AA}$, $C_6 = 49.4 \, \text{eV} \, {\text{\AA}}^6$. From (\ref{Rrep}) we then get $R_{Rep} \simeq 1.52 \, \text{\AA}$. Since the radius of the Lithium atom, $ \simeq1.45 \, \text{\AA}$, is quite smaller than the alkali-parahydrogen distance in the solid matrix, $a^{Li-p(H_2)}=3.78 \, \text{\AA}$, our pairwise-summation of the potentials between nearest-neighbors pairs, as well as neglecting higher multipolar dispersion interactions, should be reasonably good approximations.
Since the Li-H\textsubscript{2} distance in the solid matrix is smaller than the equilibrium distance, $4.76 \, \text{\AA}$, we expect that the Morse potential should give a better approximation to the true potential than the Buckingham-Hill potential.

With the numerical values above, taking the second derivative of the Buckingham-Hill potential (\ref{eq:Buckingham-Hill1}) (Eq. (\ref{curvature})), and evaluating it for the distance $r = \simeq 3.78 \text{\AA}$ between the alkali atom and any of the nearest neighbors parahydrogen molecules (see Fig. \ref{Fig2}), we obtain
$\lvert \kappa \rvert \simeq 4.34 \cdot 10^{-3} \, \text{eV} {\text{\AA}}^{-2}$, and then $\alpha = \sqrt{\lvert k \rvert/(2D)} \simeq 0.746 \, {\text{\AA}}^{-1}$, to be used in the Morse potential (\ref{Morse}).

We thus obtain
\begin{equation}
\label{Li-H2}
V_{\text{M}}^{Li-H_2}(r=3.78 \, \text{\AA}) \simeq 0.625 \cdot 10^{-3} \, \text{eV}.
\end{equation}
Taking into account that each Lithium atom has 6 nearest neighbors parahydrogen molecules of the solid matrix, and assuming a pairwise summation, we obtain the total interaction energy of each Litium atom in the parahydrogen solid matrix
\begin{equation}
\label{EnergyLi-H2}
V_{tot}^{Li-H_2}\simeq 3.7 \cdot 10^{-3} \, \text{eV} .
\end{equation}

We can also calculate the Li-(p-H\textsubscript{2}) interaction energy using the potential (\ref{Buck}), with the following parameters given in \cite{Cheng-Whaley96}; in atomic units, for the $X\Sigma$ configuration (ground state), they are: $\alpha = -2.976$, $\beta = 0.1623$, $\gamma = 0.05033$, $r_m=9.856$, $C_6=85.00$, $C_8=4050$, $C_{10}=0$ \cite{Cheng-Whaley96}. We obtain
\begin{equation}
\label{Li-H2-B}
V_{\text{B}}^{Li-H_2}(r=3.78 \, \text{\AA}) \simeq 0.147 \cdot 10^{-3} \, \text{eV}.
\end{equation}
This value is about one half of that predicted from the Morse potential, and so it is even more favorable for the feasibility of our proposed detection scheme. In fact, after summation over the six nearest neighbors parahydrogen molecules, it yields a total potential energy of the order of one meV.

Comparing (\ref{Li-H2}) and (\ref{Li-H2-B}) with (\ref{potential:Rb-Ar}) and (\ref{potential:Rb-Ne}), we see that the interaction energy in the present case is quite smaller than for both Rb-Ar and Rb-Ne. On the basis of our previous considerations, this should give a quite smaller energy width of the Lithium ground state in the matrix, at least for the contribution from its interaction with the lattice molecules, and assuming a similar dispersion of the positions of the alkali atoms in the cold solid matrix. Some experimental spectroscopic data on the Li-(p-H\textsubscript{2}) system can be found in \cite{Fajardo93},
while theoretical aspects relevant for the absorption spectra are given in \cite{Cheng-Whaley96}. In \cite{Fajardo93}, the spectral linewidth of the Lithium atoms in the p-H\textsubscript{2} cold matrix are of the order of 100 meV, but they cannot be directly compared with the ground-state interaction energies here evaluated. In fact, spectral lines, as already mentioned for the Ar and Ne cases, involve also other effects, such as, for example, the width of the excited level, energy spitting of degenerate excited levels due to crystal field, as well as phonon excitation in the crystal, that could be relevant since parahydrogen is a quantum crystal. We however expect that these effects play a less relevant role in our case, when we consider the transition between the alkali-atom Zeeman sublevel $b$ (reached after the assumed axion absorption) to the crystal conduction band, whose edge in Fig. \ref{Fig1} is indicated by $c$. The reason is that in this case, the atomic wavefunction of the sublevels $a$ and $b$ is the same, and there are not other discrete levels involved in the transition.

Finally, taking into account the theoretical results for Li-(p-H\textsubscript{2}) of this section, we expect that the Lithium-parahydrogen system could be a very promising physical system for the proposed axion detecting scheme, exploiting the transitions in Fig. \ref{Fig1} with $c$ indicating the edge of the conduction band of the cold crystal.

\section{Conclusion}
\label{sec4}
We have considered the spectroscopy of a solid crystal of inert elements, specifically Argon, Neon, and parahydrogen, at cryogenic temperatures, doped with dilute alkali atoms.
We have reported experimental spectra of dilute Rubidium atoms in a solid matrix of Argon and Neon, and analyzed the observed line widths. Then, using a simple theoretical model and known tabulated data, we have estimated the interaction energy of a single Rubidium atom with the solid Argon or Neon matrix, and, after comparison with the measured widths of the spectral lines, estimated the inhomogenous spreading of the guest atoms position inside the solid matrix.
We have also theoretically estimated the ground-state interaction energies between the dilute doping atoms with the cold matrix using simple and flexible analytical models, specifically for Rb-Ar, Rb-Ne and Lithium atoms in a solid parahydrogen matrix. We have found that the interaction energy for Li-(p-H\textsubscript{2}) is quite smaller than in Rb-Ar and Rb-Ne. This indicates that a sharper linewidth of the ground state of the doping atom, due to its interaction with the matrix, is expected in this last case. As discussed in the paper, this work is mainly aimed to a feasibility study of the recently proposed detection scheme for cosmological axions in the meV range (and also other searches for physics beyond the Standard Model), exploiting the Matrix Isolation Spectroscopy technique of magnetic-type transitions between Zeeman sublevels of alkali atoms in a cold solid matrix, induced by the axion absorption. For this proposed detection scheme, the linewidth of the alkali atoms should be reduced as much as possible. We have concluded that Lithium in a parahydrogen matrix systems could be a promising candidate for the detection scheme analyzed in this paper.

\begin{acknowledgments}
This work was supported by the INFN Scientific Committee 5, Italy, under the AXIOMA, DEMIURGOS and PHYDES projects. GF, AN, RP and LR gratefully acknowledge financial support from the Julian Schwinger Foundation and MUR. The authors wish to thank Prof. A. Stolyarov for the communication of relevant data from their recent \textit{ab initio} interatomic potential calculations. The precious technical work of E. Berto, F. Calaon, M. Rebeschini, M. Tessaro and M. Zago is also acknowledged.
\end{acknowledgments}

\bibliography{biblio1}

\end{document}